# 3×3 transfer matrix modelling

Matteo Cherchi, VTT – Technical Research Centre of Finland

Unlike common devices based on ring resonators, the structure in Fig. 1.a involves not only 2×2 couplers but also a 3×3 coupler, which means that a 3×3 transfer matrix approach is required to model the system. To the best of our knowledge, no such a model has been developed before. The only model available in the literature is based on a clever recursive 2×2 transfer matrix model [1], which requires lengthy calculations that depend on the chosen boundary conditions and on the particular geometry chosen. The scope of this document is instead to show how to generalize the standard 2×2 transfer matrix approach to cover any system with 3×3 couplers, and calculate the transfer matrix of any complicated system just as a product of simple 3×3 matrices.

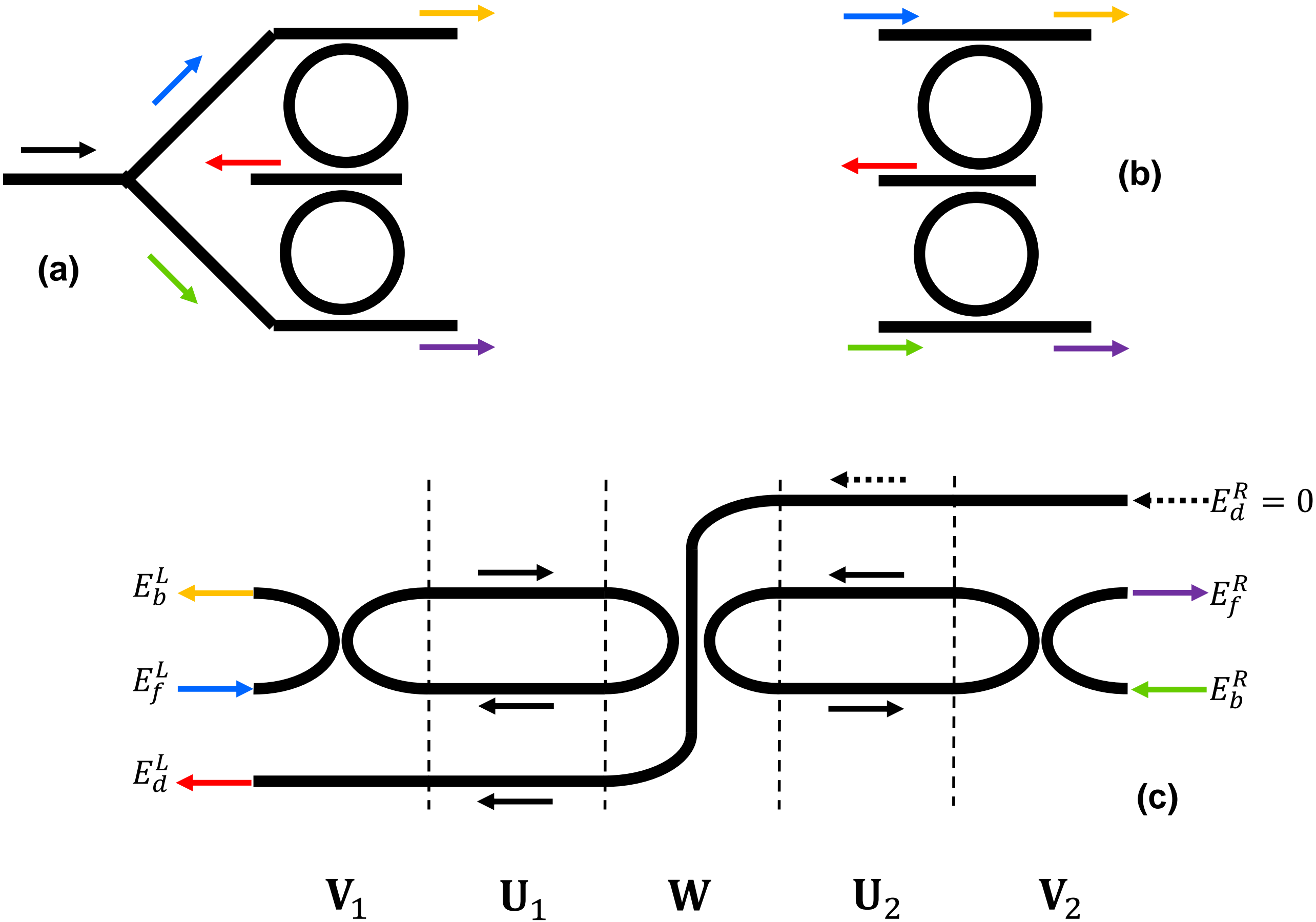

*Fig. 1 (a) Schematic of the structure under study, (b) relevant section to be modelled, and (c) topologically equivalent circuit divided in 5 sections to build a suitable transfer matrix model.*

We start by noticing that the relevant part of the system to be modelled is simply the one in Fig. 1(b), being the double input just a matter of boundary conditions. We then suitably divide the system in a topologically equivalent cascade of 3-arm sections, including two sections with 2×2 couplers and one uncoupled waveguide, two sections with three uncoupled arms, and one section where all three waveguides are coupled to each other, as shown in Fig. 1.c. All sections are characterized by a forward propagating field $E_f$ and two backward propagating fields $E_b$ and $E_d$. In particular we want to determine the transfer matrix $\mathbf{T}$ linking the field on the right-hand side to the left-hand side of the system, such that

$$\psi_R = \mathbf{T}_{\rm tot}\psi_L \ , \tag{1}$$

where the vectors $\psi_X$ $(X = R, L)$ are defined as

$$\psi_X = \begin{pmatrix} E_f^X \\ E_b^X \\ E_d^X \end{pmatrix}. \tag{2}$$

From the different sections in Fig. 1.c, we can see that the transfer matrix $\mathbf{T}$ is in fact a product of five matrices:

$$\mathbf{T}_{\rm tot} = \mathbf{V}_2\mathbf{U}_2\mathbf{W}\mathbf{U}_1\mathbf{V}_1 \ , \tag{3}$$

where the $\mathbf{V}_p$ matrices account for the 2×2 coupling, the $\mathbf{U}_p$ propagate the uncoupled waveguides and the matrix $\mathbf{W}$ models the 3×3 coupling.

Matrices $\mathbf{U}_p$ can be easily inferred from the 2×2 model, and can be written, without any loss of generality, as

$$\mathbf{U}_p = \begin{pmatrix} e^{i\varphi_p} & 0 & 0 \\ 0 & 1/e^{i\varphi_p} & 0 \\ 0 & 0 & 1 \end{pmatrix}, \tag{4}$$

where the phase change of $E_d$ can be arbitrarily set to zero, while $\varphi_p \equiv k_p L_p/2$ are the phases accumulated by the $E_f$ and $E_b$ fields propagating with propagation constant $k_p$ (that can have also non-negligible imaginary part, to account for propagation losses) in half the ring length $L_p$.

Similarly the matrices $\mathbf{V}_p$ can be written as

$$\mathbf{V}_p = \frac{1}{it_p}\begin{pmatrix} -1 & r_p^* & 0 \\ -r_p & 1 & 0 \\ 0 & 0 & it_p \end{pmatrix}, \tag{5}$$

where the element $(\mathbf{V}_p)_{33} = 1$ imposes again zero phase change to $E_d$, without any loss of generality (the asterisk stands for complex conjugate). The top-left 2×2 submatrix is nothing but the standard transfer matrix of a coupler [2–4] (or, equivalently, of a mirror) with field transmission $t_p$ and field reflection $r_p$.

The derivation of matrix $\mathbf{W}$ requires instead some more effort. It is actually instructive to recall how the transfer matrix of the 2×2 coupler is derived from the scattering matrix [5] in the 2×2 case, so to follow a similar approach for the 3×3 case. In the 2×2 case the scattering matrix $\mathbf{S} = (s_{ij})$ links input and output of the coupler (see Fig. 2.a) as follows:

$$\phi_O = \mathbf{S}\phi_I \ , \tag{6}$$

where input and output vectors $\phi_X$ $(X = I, O)$ are defined as

$$\phi_X = \begin{pmatrix} E_1^X \\ E_2^X \end{pmatrix}. \tag{7}$$

We want to derive the 2×2 transfer matrix $\mathbf{T}$ linking left- and right-hand side of the coupler (Fig. 2.b) as follows

$$\psi_R = \mathbf{T}\psi_L \ , \tag{8}$$

where the vectors $\psi_X$ $(X = R, L)$ are defined as

$$\psi_X = \begin{pmatrix} E_f^X \\ E_b^X \end{pmatrix}. \tag{9}$$

From the linear system of equations corresponding to Eq. (6), using the identities $E_f^L = E_1^I$, $E_b^L = E_1^O$, $E_f^R = E_2^O$, and $E_b^R = E_2^I$, it is straightforward to calculate $\mathbf{T}$ as

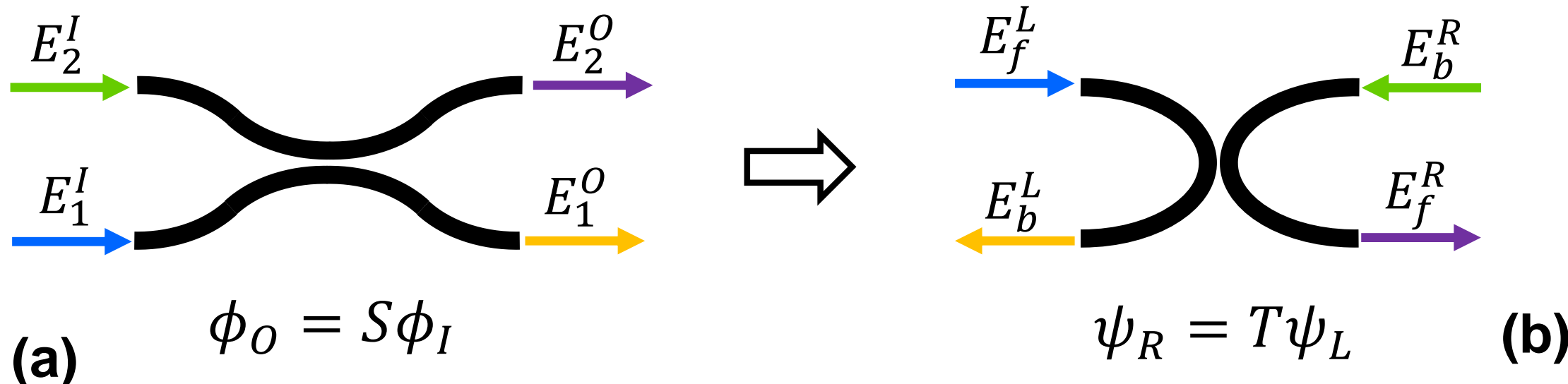

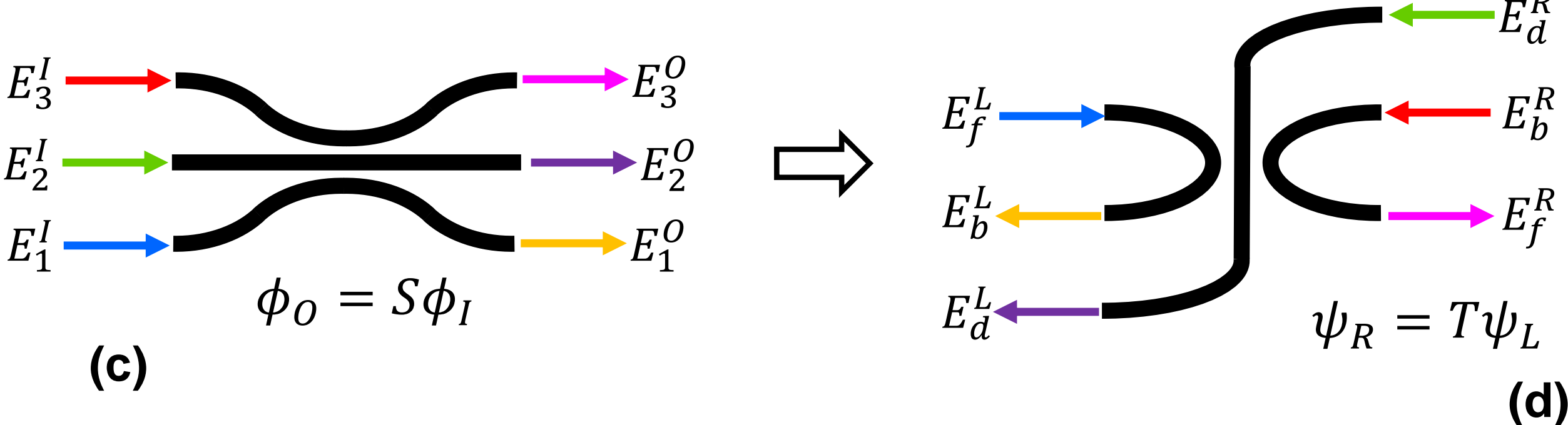

*Fig. 2 (a) Scattering matrix model of a 2×2 coupler, (b) transfer matrix model for the same coupler, (c) scattering matrix model of a 3×3 coupler, and (d) transfer matrix model for the same coupler.*

$$\mathbf{T} = \frac{1}{s_{12}} \begin{pmatrix} |\mathbf{S}| & s_{22} \\ -s_{11} & 1 \end{pmatrix} , \tag{10}$$

where $|\mathbf{S}|$ is the determinant of the matrix $\mathbf{S}$. Given the standard form of the scattering matrix $\mathbf{S}$:

$$\mathbf{S} = \begin{pmatrix} r & it \\ it & r^* \end{pmatrix} , \tag{11}$$

and assuming a lossless coupler ($|\mathbf{S}| = 1$), leads to

$$\mathbf{T} = \frac{1}{it} \begin{pmatrix} -1 & r^* \\ -r & 1 \end{pmatrix} . \tag{12}$$

$\mathbf{T}$ is also unitary, with $|\mathbf{T}| = 1$. For synchronous couplers:

$$\begin{cases} r = \cos(\kappa L_c) \\ t = \sin(\kappa L_c) \end{cases}, \tag{13}$$

where $\kappa$ is the coupling coefficient and $L_c$ is the effective length of the coupler. We can also extend these relations to asynchronous couplers. Assuming waveguides with propagation constant differing by $\delta$, reflection and transmission coefficient can be written as follows [6]:

$$\begin{cases} r = \cos(\mu L_c) + i\sin(\mu L_c)\cos(\gamma) \\ t = \sin(\mu L_c)\sin(\gamma) \end{cases}, \tag{14}$$

where $\mu \equiv \sqrt{\kappa^2 + (\delta/2)^2}$ and $\cos\gamma \equiv -\delta/(2\mu)$ (which implies $\sin\gamma \equiv \kappa/\mu$).

We want now to apply the same strategy to derive the 3×3 transfer matrix $\mathbf{T}$ of the tri-coupler in Fig. 2.d from the 3×3 scattering matrix $\mathbf{S}$ of Fig. 2.c. In this case the following identities hold: $E_f^L = E_1^I$, $E_b^L = E_1^O$, $E_d^L = E_2^O$, $E_f^R = E_3^O$, $E_b^R = E_3^I$, and $E_d^R = E_2^I$, leading to

$$\mathbf{T} = \frac{1}{M_{31}} \begin{pmatrix} |S| & -M_{13} & M_{21} \\ M_{33} & -s_{22} & s_{12} \\ -M_{32} & s_{23} & -s_{13} \end{pmatrix}, \tag{15}$$

where $M_{ij} \equiv \left|(s_{pq})_{p\neq i, q\neq j}\right|$ are the so called (i,j)minors of the matrix $\mathbf{S}$, i.e. the determinants of the submatrices obtained by eliminating the *i*-th row and the *j*-th column.

At this point, we just miss the scattering matrix of the coupler in Fig. 2.c, which we could not find anywhere in the literature. To calculate the matrix we follow the same approach used in reference [6], i.e. starting from the differential equations of the coupled mode theory in space [7,8], found in the existing literature on tri-couplers [9,10]. We will focus on asynchronous couplers where the two external waveguides are identical, whereas the middle waveguide is different, being $\delta$ the difference of propagation constants. Therefore, assuming nearest neighbour coupling only and no losses, the system of differential equation is simply:

$$\begin{cases} \dfrac{dE_1}{dz} = i\kappa E_2 + i\dfrac{\delta}{2}E_1 \\ \dfrac{dE_2}{dz} = i\kappa(E_1 + E_3) - i\dfrac{\delta}{2}E_2 \\ \dfrac{dE_3}{dz} = i\kappa E_2 + i\dfrac{\delta}{2}E_3 \end{cases}, \tag{16}$$

and by defining the symmetric and anti-symmetric combinations $E_S \equiv \frac{1}{\sqrt{2}}(E_1 + E_3)$ and $E_A \equiv \frac{1}{\sqrt{2}}(E_1 - E_3)$ they reduce two the 2×2 case, being $E_A$ already an eigenmode of the system:

$$\begin{cases} \dfrac{dE_2}{dz} = i\sqrt{2}\kappa E_S - i\dfrac{\delta}{2}E_2 \\ \dfrac{dE_S}{dz} = i\sqrt{2}\kappa E_2 + i\dfrac{\delta}{2}E_S \\ \dfrac{dE_A}{dz} = 0 \end{cases}. \tag{17}$$

Here $\kappa$ denotes the coupling coefficient between two adjacent waveguides. If we introduce the input and output vectors $\Phi_X$ $(X = I, O)$

$$\Phi_X = \begin{pmatrix} E_2^X \\ E_S^X \\ E_A^X \end{pmatrix}, \tag{18}$$

the solution can be written as $\Phi_O = \bar{\mathbf{S}}\Phi_I$ [6], where $\bar{S}$ is the scattering matrix in the rotated basis $\{E_2, E_S, E_A\}$

$$\bar{\mathbf{S}} \equiv \begin{pmatrix} \varrho & i\tau & 0 \\ i\tau & \varrho^* & 0 \\ 0 & 0 & 1 \end{pmatrix}. \tag{19}$$

and we have defined the reflection and transmission coefficients

$$\begin{cases} \varrho = \cos(\mu L_c) + i\sin(\mu L_c)\cos(\gamma) \\ \tau = \sin(\mu L_c)\sin(\gamma) \end{cases}, \tag{20}$$

being $L_c$ the effective length of the couplers, $\mu \equiv \sqrt{2\kappa^2 + (\delta/2)^2}$ and $\cos\gamma \equiv -\delta/(2\mu)$ (which implies $\sin\gamma \equiv \sqrt{2}\kappa/\mu$). Synchronous couplers are just the limiting case for $\delta = 0$, leading to $\varrho = \cos(\sqrt{2}\kappa L_c)$ and $\tau = \sin(\sqrt{2}\kappa L_c)$.

Going back in the original basis $\{E_1, E_2, E_3\}$ can be easily done through the matrix of basis change

$$\mathbf{R} = \frac{1}{\sqrt{2}} \begin{pmatrix} 1 & 0 & 1 \\ 0 & \sqrt{2} & 0 \\ 1 & 0 & -1 \end{pmatrix}, \tag{21}$$

which happens to be an involutory matrix, i.e. such that $\mathbf{R} = \mathbf{R}^{-1}$. The scattering matrix $\mathbf{S}$ takes the form

$$\mathbf{S} = \mathbf{R}\bar{\mathbf{S}}\mathbf{R} = \frac{1}{2} \begin{pmatrix} 1+\varrho & i\sqrt{2}\tau & -(1-\varrho) \\ i\sqrt{2}\tau & 2\varrho^* & i\sqrt{2}\tau \\ -(1-\varrho) & i\sqrt{2}\tau & 1+\varrho \end{pmatrix}, \tag{22}$$

which is the unitary matrix describing the coupler of Fig. 2.c. From Eq. (15) we can finally write the transfer matrix for Fig. 2.d, that is the missing matrix $\mathbf{W}$ from Fig. 1.c:

$$\mathbf{W} = \frac{1}{1-\varrho^*} \begin{pmatrix} -2 & 1+\varrho^* & -i\sqrt{2}\tau \\ -(1+\varrho^*) & 2\varrho^* & -i\sqrt{2}\tau \\ i\sqrt{2}\tau & -i\sqrt{2}\tau & -(1-\varrho) \end{pmatrix}. \tag{23}$$

Noticeably, $\mathbf{W}$ is unitary, with $|\mathbf{W}| = -(1-\varrho)/(1-\varrho^*)$. This result completes the 3×3 transfer matrix model of the coupled ring resonators.

One last step is still necessary to impose the boundary conditions of Fig. 1.a and calculate the unknowns $E_b^L$, $E_d^L$, and $E_f^R$, assuming input fields $E_f^L = 1/\sqrt{2}$ and $E_f^L = e^{i\vartheta}/\sqrt{2}$, where for the sake of generality we are assuming a relative phase difference between the two inputs. By explicitly writing the system of linear equations $\psi_R = \mathbf{T}_{\text{tot}}\psi_L$, in terms of the matrix components $\mathbf{T}_{\text{tot}} = (t_{ij})$ and of the fields components, it can be easily calculated:

$$\psi_L = \begin{pmatrix} \frac{1}{\sqrt{2}} \\ E_b^L \\ E_d^L \end{pmatrix}, \psi_R = \begin{pmatrix} E_f^R \\ \frac{e^{i\vartheta}}{\sqrt{2}} \\ 0 \end{pmatrix} \quad \Rightarrow \quad \begin{cases} E_b^L = \dfrac{e^{i\vartheta} t_{33} - M_{12}}{\sqrt{2} M_{11}} \\ E_d^L = -\dfrac{e^{i\vartheta} t_{32} - M_{13}}{\sqrt{2} M_{11}} \\ E_f^R = \dfrac{e^{i\vartheta} M_{21} + |T_{\text{tot}}|}{\sqrt{2} M_{11}} \end{cases}, \tag{24}$$

where $M_{ij} \equiv \left|(t_{pq})_{p\neq i, q\neq j}\right|$ indicate once again the (i,j)minors of the matrix $\mathbf{T}_{\text{tot}}$.

The same way it is possible to treat any other boundary conditions. For example the response of the system when light is launched in one input only can be easily calculated as:

$$\psi_L = \begin{pmatrix} 1 \\ E_b^L \\ E_d^L \end{pmatrix}, \psi_R = \begin{pmatrix} E_f^R \\ 0 \\ 0 \end{pmatrix} \quad \Rightarrow \quad \begin{cases} E_b^L = -\dfrac{M_{12}}{M_{11}} \\ E_d^L = \dfrac{M_{13}}{M_{11}} \\ E_f^R = \dfrac{|T_{\text{tot}}|}{M_{11}} \end{cases}. \tag{25}$$

## References

1. L. Zhou, R. Soref, and J. Chen, "Wavelength-selective switching using double-ring resonators coupled by a three-waveguide directional coupler," Opt. Express **23**, 13488–13498 (2015).
2. P. Yeh, *Optical Waves in Layered Media* (Wiley, 2004).
3. M. Born, E. Wolf, and A. B. Bhatia, *Principles of Optics: Electromagnetic Theory of Propagation, Interference and Diffraction of Light* (Cambridge University Press, 1999).
4. M. Cherchi, "Bloch analysis of finite periodic microring chains," Appl. Phys. B **80**, 109–113 (2004).
5. H. A. Haus, *Waves and Fields in Optoelectronics* (Prentice Hall, Incorporated, 1984).
6. M. Cherchi, "Wavelength-Flattened Directional Couplers: A Geometrical Approach," Appl. Opt. **42**, 7141 (2003).
7. H. A. Haus and W. Huang, "Coupled-mode theory," Proc. IEEE **79**, 1505–1518 (1991).
8. W.-P. Huang, "Coupled-mode theory for optical waveguides: an overview," J. Opt. Soc. Am. A **11**, 963–983 (1994).
9. K. Iwasaki, S. Kurazono, and K. Itakura, "The coupling of modes in three dielectric slab waveguides," Electron. Commun. Jpn. **58**, 100–108 (1975).
10. H. Haus and C. Fonstad, "Three-waveguide couplers for improved sampling and filtering," IEEE J. Quantum Electron. **17**, 2321–2325 (1981).